\documentclass[12pt]{iopart}

\usepackage{graphicx}

\newcommand{\pT}{p_\mathrm{T}}
\newcommand{\pTass}{p_\mathrm{T}^{assoc}}
\newcommand{\kt}{k_\mathrm{T}}
\newcommand{\akt}{\mathrm{anti-}k_\mathrm{T}}

\newcommand{\gev}{\mathrm{GeV}}
\newcommand{\gevc}{\mathrm{GeV}/c}
\newcommand{\raa}{R_\mathrm{AA}}
\newcommand{\rcp}{R_\mathrm{CP}}
\newcommand{\iaa}{I_\mathrm{AA}}
\newcommand{\daa}{D_\mathrm{AA}}
\newcommand{\flow}{v_{2}}
\newcommand{\fSTAR}{for the STAR Collaboration}

\begin{document}

% Usage: \title[Short title]{Full title}
% [Short title] is optional; use where title is too long
% or contains footnotes, 50 characters maximum 
\title[Jet studies with STAR at RHIC]{Jet studies with STAR at RHIC: jet algorithms, jet shapes, jets in AA}

\author{J Kapit\'an (for the STAR Collaboration)}

\address{Na Truhlarce 38/64, Praha 8, Czech Republic}
\ead{kapitan@rcf.rhic.bnl.gov}

\begin{abstract}
Hard scattered partons are predicted to be well calibrated probes of the hot and dense medium produced in heavy ion collisions. Interactions of these partons with the medium will result in modifications of internal jet structure in Au+Au events compared to that observed in the p+p/d+Au reference. Full jet reconstruction is a promising tool to measure these effects without the significant biases present in measurements with high-$\pT$ hadrons.

One of the most significant challenges for jet reconstruction in the heavy ion environment comes from the correct characterization of the background fluctuations. The jet momentum irresolution due to background fluctuations has to be  understood in order to recover the correct jet spectrum. Recent progress in jet reconstruction methodology is discussed, as well as recent measurements from p+p, d+Au and Au+Au collisions at $\sqrt{s_\mathrm{NN}}=200~\gev$.

\end{abstract}

\section{Introduction}

Jets are remnants of hard-scattered partons, which are the fundamental objects of perturbative QCD. At Relativistic Heavy Ion Collider (RHIC), they can be used as a probe of the hot and dense matter created in heavy ion collisions. Interaction and energy loss of energetic partons in the medium lead to jet quenching in heavy ion collisions. Until recently, jet quenching was studied indirectly using single particle spectra and di-hadron correlations~\cite{quenching}. These measurements are however limited in the sensitivity to probe partonic energy loss mechanisms due to biases toward hard fragmentation and small energy loss~\cite{trenk}.

Developments in theory (for example~\cite{fj,bgsub}) and experiment (detector upgrades, increased RHIC luminosity) finally enabled full jet reconstruction in heavy ion collisions~\cite{initial}. Full jet reconstruction reduces the biases of indirect measurements and enables access to qualitatively new observables such as energy flow and fragmentation functions. As a baseline measurement for heavy ion jet studies, p+p collisions at the same energy are used. To isolate initial state effects from medium modification, measurements in d+Au are essential.

We present current jet analyses at STAR, starting with recent results on initial state effects (d+Au). Status of jet spectra analysis in Au+Au follows, including studies of background fluctuations and their effect on the measurement of jet spectra. Finally we discuss recent results on correlations in Au+Au triggered by fully reconstructed jets: di-jet analysis and jet-hadron correlations.

\section{Jet reconstruction}

The present analysis is based on $\sqrt{s_\mathrm{NN}} = 200~\gev$ data from the STAR experiment, recorded during 2006-2008. The Barrel Electromagnetic Calorimeter (BEMC) detector is used to measure the neutral component of jets, and the Time Projection Chamber (TPC) detector is used to measure the charged particle component of jets. In the case of a TPC track pointing to a BEMC tower, its momentum is subtracted from the tower energy to avoid double counting (electrons, MIP and possible hadron showers in the BEMC). Pseudorapidity acceptance for jets is $|\eta| < 0.6$ in p+p and Au+Au and $|\eta| < 0.55$ in the case of d+Au collisions.

Recombination jet algorithms $\kt$ and $\akt$, part of the FastJet package~\cite{fj}, are used for jet reconstruction. To subtract the background, a method based on active jet areas~\cite{bgsub} is applied event-wise: $\pT^\mathrm{Rec} = \pT^\mathrm{Candidate} - \rho \cdot A$, with $\rho$ estimating the background density per event and $A$ being the jet active area. 

An important aspect of underlying event background are its fluctuations. We discuss data-driven methods used to correct the jet observables for these fluctuations.

\section{Initial state: d+Au}

This analysis is based on minimum bias triggered $\sqrt{s_\mathrm{NN}} = 200~\gev$ data from the STAR experiment, recorded during RHIC run 8 (2007-2008). The Beam Beam Counter detector, located in the Au nucleus fragmentation region, was used to select the 20\% highest multiplicity events in d+Au collisions. 10M events after event cuts were used for jet finding ($\akt$ algorithm) with a resolution parameter $R = 0.4$ and $\pT > 0.2~\gevc$ cut was applied to tracks and towers. 

PYTHIA 6.410 and GEANT detector simulations (adjusted to match the realistic TPC tracking efficiency in d+Au run 8 running) were used for jet corrections to hadron level. Embedding into real d+Au events at level of reconstructed tracks and towers was used to correct for background fluctuations. A bin-by-bin correction was applied to the jet spectrum~\cite{hp2010}.

To compare the per event jet yield in d+Au to jet cross section measurements in p+p collisions, MC Glauber studies were utilized: $\langle N_\mathrm{bin} \rangle = 14.6 \pm 1.7$ for 0-20\% highest multiplicity d+Au collisions and $\sigma_\mathrm{inel,pp} = 42~\mathrm{mb}$. These factors were used to scale the p+p jet cross section measured previously by the STAR collaboration~\cite{STAR-ppJetPRL} using a Mid Point Cone (MPC) jet algorithm with a cone radius of $R = 0.4$. The resulting d+Au jet $\pT$ spectrum is shown in Figure~\ref{fig:dAuptsp} together with the scaled p+p jet spectrum. Within the systematic uncertainties, the d+Au jet spectrum scales with $\langle N_\mathrm{bin} \rangle$. 

The leading systematic uncertainty is the Jet Energy Scale (JES) that is driven by imprecise knowledge of TPC tracking efficiency for tracks in jets with realistic run 8 d+Au detector backgrounds. This will be considerably improved by embedding jets in raw d+Au data. With better handle on JES and by measuring jet spectrum in run 8 p+p collisions and in peripheral d+Au collisions, we'll be able to construct $\raa$ and $\rcp$ for jets, respectively.

\section{Inclusive jet spectra and background fluctuations in Au+Au}

\begin{figure}[htb]
\begin{minipage}[h]{0.52\textwidth}
\includegraphics[width=\textwidth]{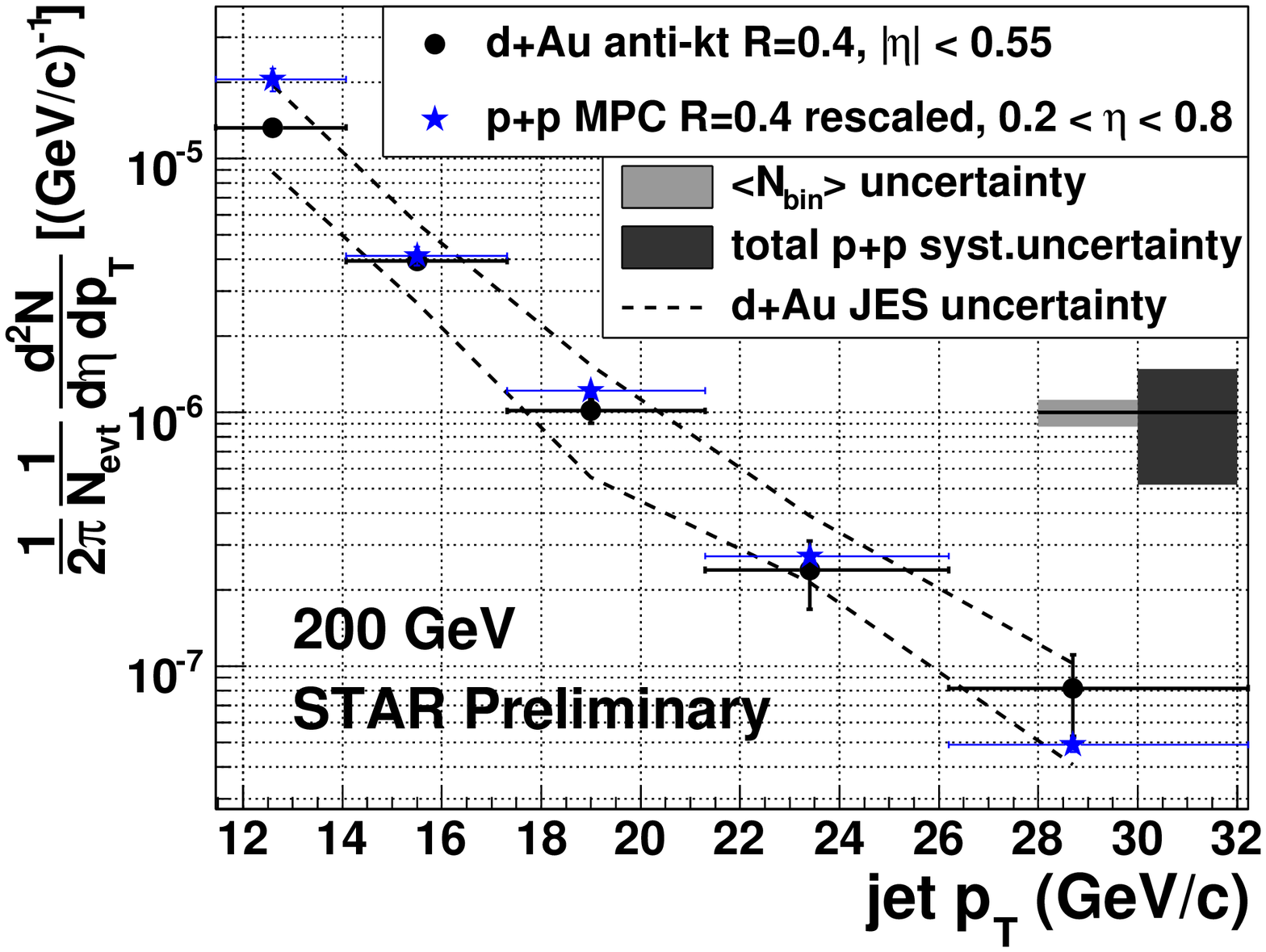}
\vspace{-11mm}
\caption{\label{fig:dAuptsp}Jet $\pT$ spectrum from d+Au collisions~\cite{hp2010} compared to $\langle N_\mathrm{bin} \rangle$ scaled p+p spectrum~\cite{STAR-ppJetPRL}.}
\end{minipage}
\hfill
\begin{minipage}[h]{0.45\textwidth}    
\includegraphics[width=\textwidth]{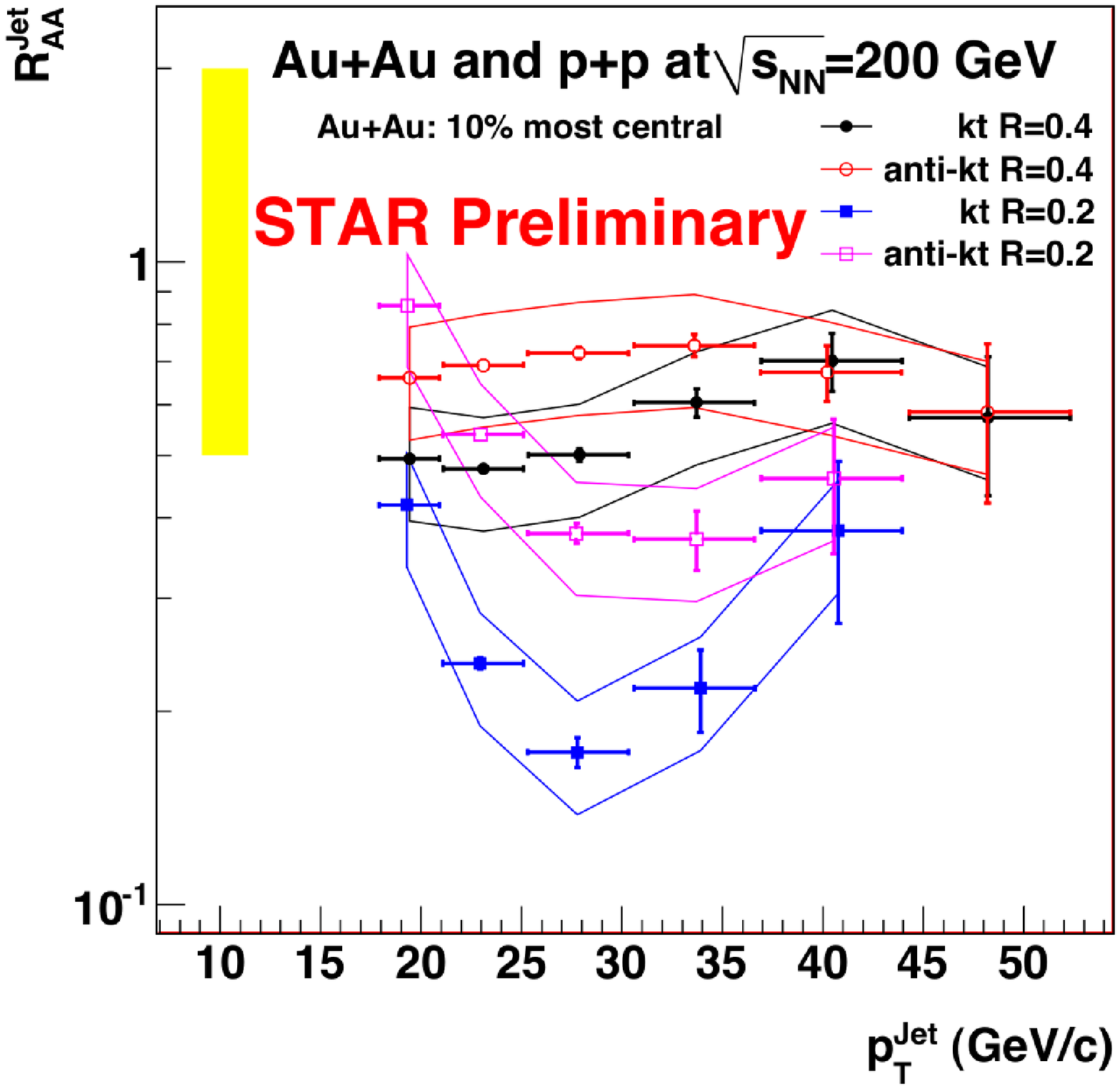}
\vspace{-9mm}
\caption{\label{fig:jetRaa}$\raa$ of jets in central Au+Au collisions for $\kt$ and $\akt$ algorithms and $R = 0.2, 0.4$~\cite{MP}.}
\end{minipage}
\end{figure}

\begin{figure}[htb]
\begin{minipage}[h]{0.425\textwidth}
\includegraphics[width=\textwidth]{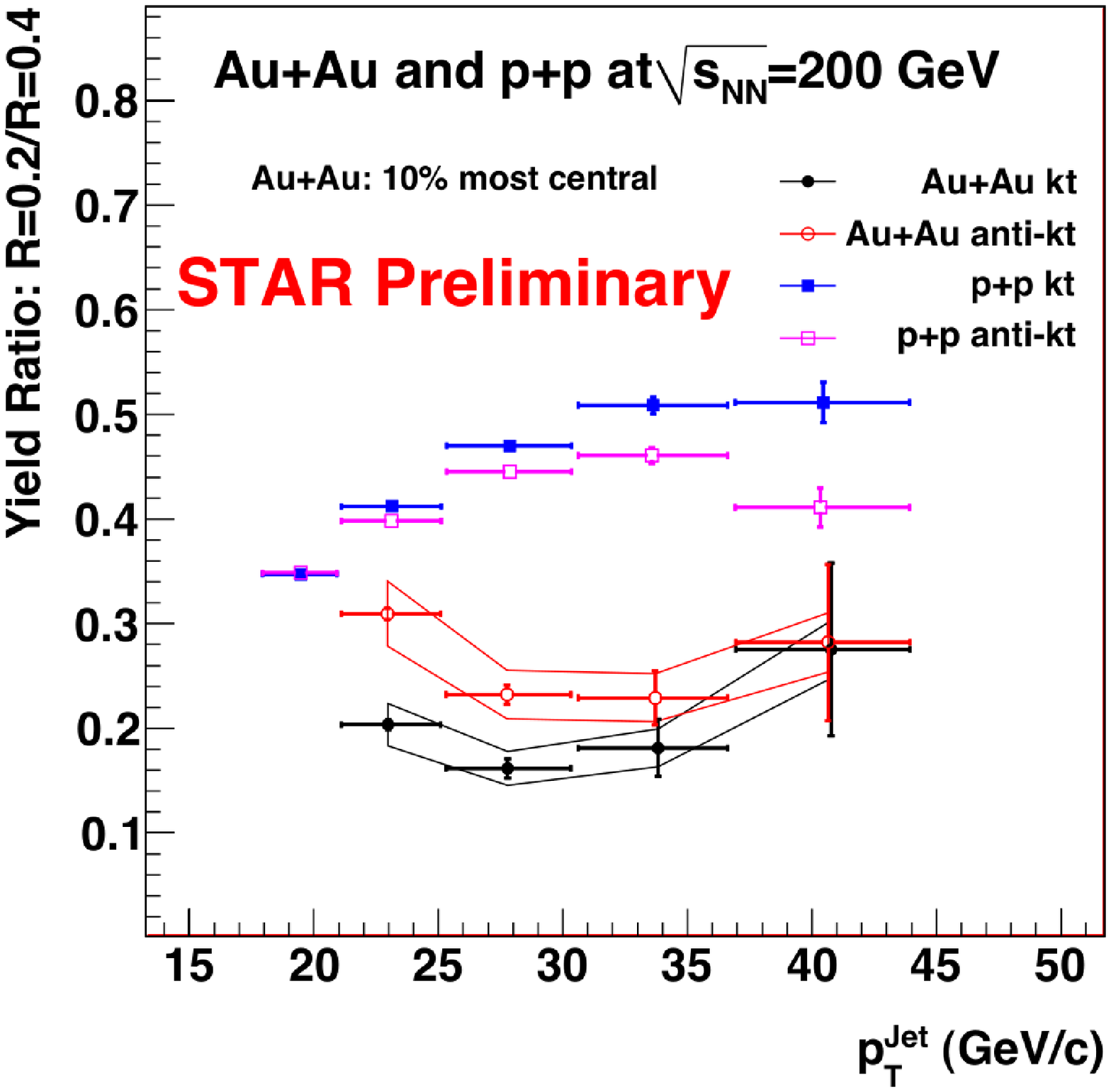}
\vspace{-10mm}
\caption{\label{fig:jetRatio}Ratio of $R=0.2/R=0.4$ jet $\pT$ spectra in p+p and Au+Au collisions~\cite{MP}.}
\end{minipage}
\hfill
\begin{minipage}[h]{0.535\textwidth}    
\includegraphics[width=\textwidth]{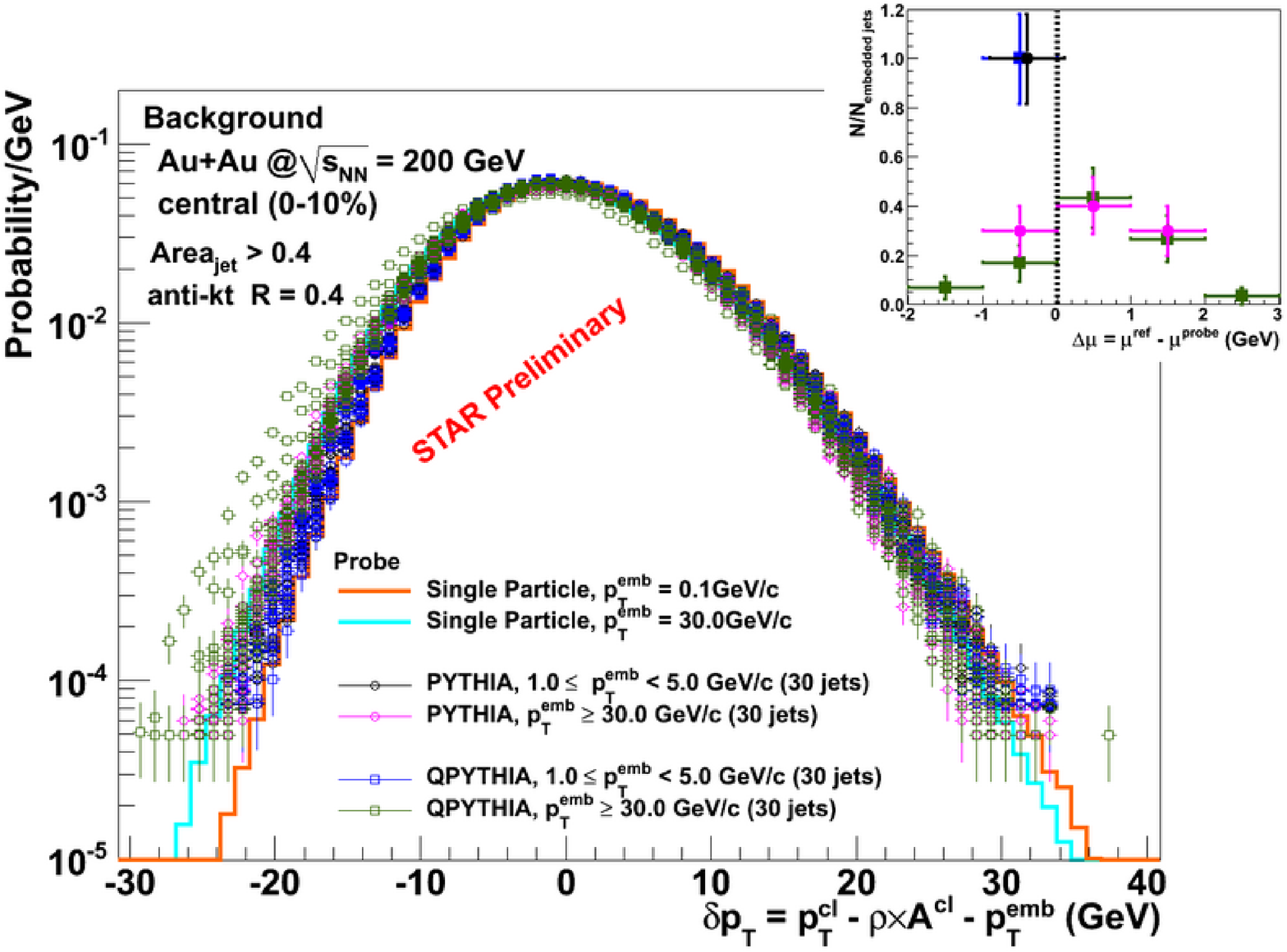}
\vspace{-10mm}
\caption{\label{fig:deltapt_fragmentation}Quantifying the background fluctuations and their dependence on probe $\pT$ and fragmentation~\cite{peter}.}
\end{minipage}
\end{figure}

Preliminary results on jet $\pT$ spectrum in Au+Au collisions at $\sqrt{s_\mathrm{NN}} = 200~\gev$ were reported in~\cite{MP}. In this analysis, the background fluctuations were estimated by generating PYTHIA jets and embedding them into real central Au+Au events. The resulting spectrum distortion was parametrized by Gaussian, for $R = 0.4$ the width is $\sigma = 6.8~\mathrm{GeV}$ with systematic uncertainty $\pm 1~\mathrm{GeV}$. This parametrization was then used for a regularized matrix inversion to unfold the measured jet spectrum. Resulting $\raa$ is shown in Figure~\ref{fig:jetRaa}. The systematic uncertainties prevent us from precisely quantifying the suppression for $R = 0.4$ jets (their reduction will be the subject of the next paragraphs). It's clear however, that these jets are less suppressed than jets with $R = 0.2$ and charged hadrons ($\raa \approx 0.2$). This is consistent with a picture of jet profile broadening from $R = 0.2$ to $R = 0.4$ in central Au+Au collisions with respect to p+p, which is illustrated by spectra ratios in Figure~\ref{fig:jetRatio}.

\vspace{-5mm}

Precise characterization of underlying event background fluctuations is essential to reduce systematic uncertainties in jet measurements. These are hence a subject of intense study, both theoreticlly~\cite{bgfluct} and experimentally. We summarize here recent results of STAR studies of background fluctuations~\cite{peter}.

To quantify the background fluctuation, a method is used where a probe ``jet'' (single particle, PYTHIA jet, QPYTHIA jet) is embedded into real central Au+Au events. This allows to extract the shape of the fluctuations (they are known to be non-Gaussian) and also to check if they are independent of the fragmentation pattern of the probe (this is essential as jet fragmentation is modified in Au+Au collisions). We embed an object of known $\pT = \pT^\mathrm{embed}$ and apply jet reconstruction on the hybrid event ($\akt$ algorithm with $R = 0.4$). We match a reconstructed jet containing more than 50\% of probe $\pT$ to the probe jet and quantify the response of the hybrid system to the embedded jet via:

\begin{equation}
\delta\pT = \pT^\mathrm{reco} - \rho\cdot A^\mathrm{reco} - \pT^\mathrm{embed},
\end{equation}

where $A^\mathrm{reco}$ is the area of the matched reconstructed jet and $\rho$ is determined prior to the embedding step. This definition is identical to Eq. (1) in \cite{bgfluct}. The normalized distribution of $\delta\pT$  is the probability distribution to find jet energy (after event-wise background correction) $\pT^{corr}=\pT^{true}+\delta\pT$. If there were no background fluctuations, $\delta\pT$ would be a delta function at zero. For very low $\pT$ probes, areas of $\akt$ jets get very small, so a cut $A^\mathrm{reco} > 0.4$ was applied. With this cut, $\delta\pT$ distribution turns out to be largely independent of $\pT^\mathrm{embed}$~\cite{peter}.

We have investigated dependence of $\delta\pT$ on jet fragmentation pattern. Figure~\ref{fig:deltapt_fragmentation} shows the overlay of multiple $\delta\pT$ distributions for single particle jets and for jets with both low and high $\pT$ generated by PYTHIA and Q-PYTHIA ($\hat{q}=5~\mathrm{GeV}^{2}/\mathrm{fm}$). In order to compare their shapes directly, the distributions were aligned horizontally by fitting a Gaussian function to $\delta\pT<0$ and aligning the centroids by shifting relative to one reference distribution. The shifts are shown in the insert and are typically smaller in magnitude than 1 GeV. The overlay shows that the $\delta\pT$ distribution is to a large extent universal, within a factor $\sim2$ at $\delta\pT = 30~\gev$, especially in region $\delta\pT > 0$ which drives the smearing of the inclusive jet spectrum. Further quantification of this observation and its application to deconvolution of the measured inclusive jet spectrum in central Au+Au collisions is in progress.

\vspace{-5mm}

\section{Jet triggered correlations}

A highly biased jet population was used as {\it trigger} in di-jet and jet-hadron correlations. Trigger jets are required to contain a BEMC tower with $E_\mathrm{T} > 5.4~\gev$ to achieve a longer in-medium pathlength on the recoil (away side). To limit the effect of background fluctuations, trigger jets are reconstructed using only TPC tracks and BEMC towers with $\pT > 2~\gevc$. A $2~\gev$ systematic uncertainty on trigger jet energy was used to account for any remaining effect. 

\begin{figure}[htb]
\begin{minipage}[h]{0.485\textwidth}
\includegraphics[width=\textwidth]{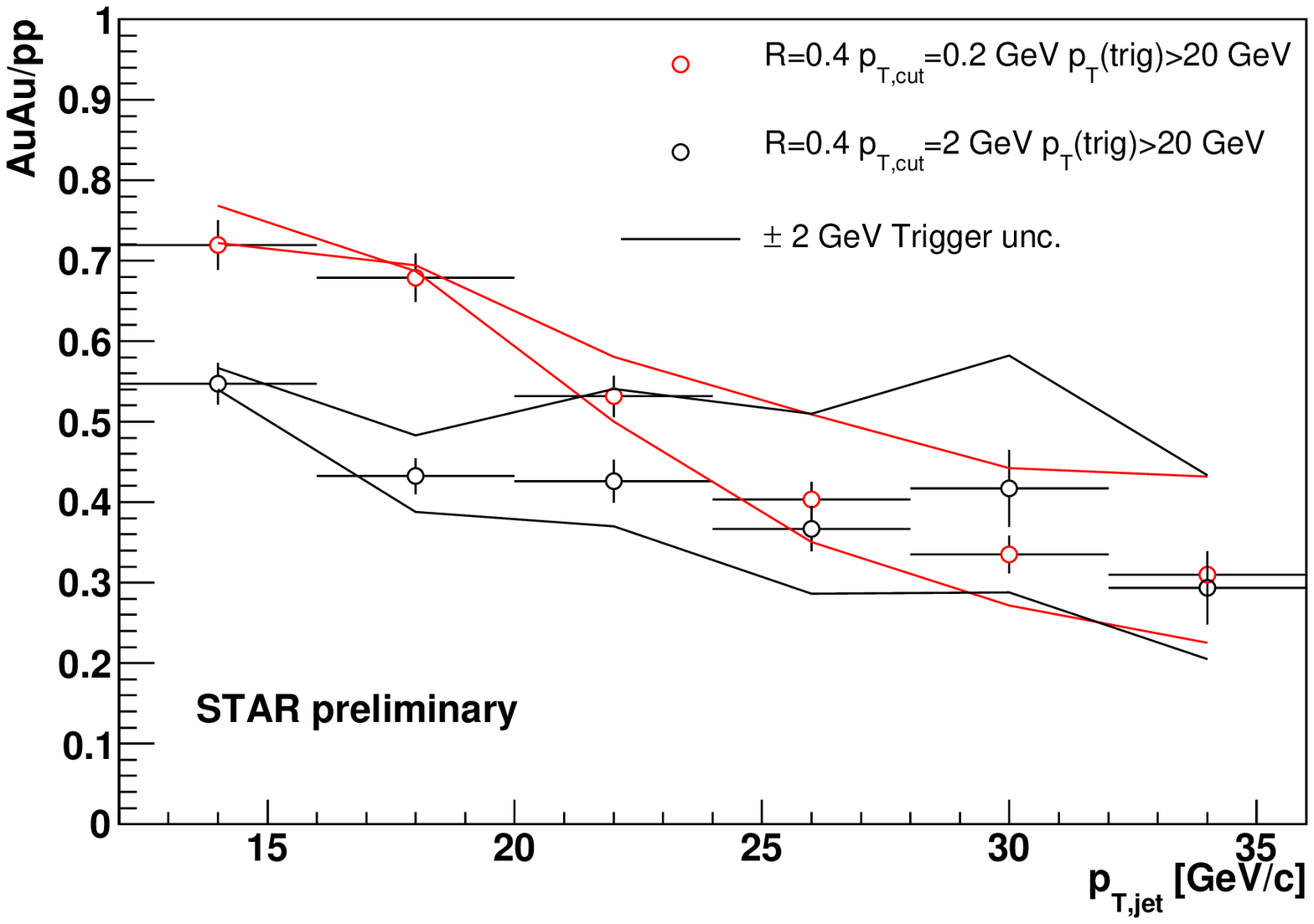}
\vspace{-10mm}
\caption{\label{fig:dijets}Ratio of recoil jet $\pT$ spectra in Au+Au/p+p~\cite{Elena_Prague}.}
\end{minipage}
\hfill
\begin{minipage}[h]{0.475\textwidth}    
\includegraphics[width=\textwidth]{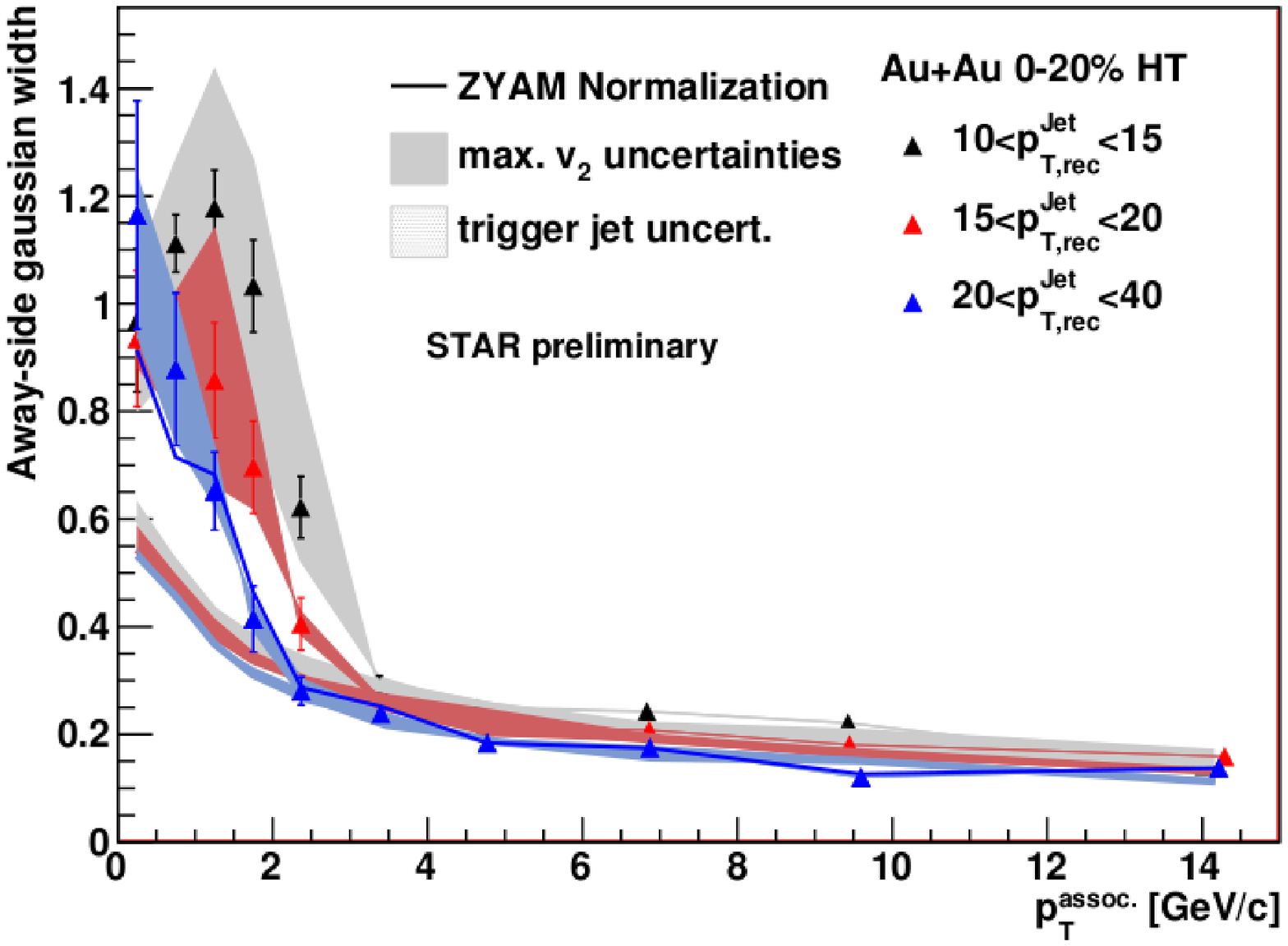}
\vspace{-10mm}
\caption{\label{fig:ASwidth}Away side Gaussian width in JH correlations~\cite{joern}.}
\end{minipage}
\end{figure}

Recoil jet $\pT$ spectrum was measured in p+p and Au+Au (0-20\% most central) collisions~\cite{Elena_Prague}. A Gaussian model of background fluctuations was used to unfold the Au+Au spectrum with systematic uncertainty $\pm 1~\mathrm{GeV}$. Figure~\ref{fig:dijets} shows a significant suppression of recoil jet $\pT$ spectrum in Au+Au compared to p+p for $R = 0.4$, which suggests jet broadening beyond $R = 0.4$. Considering also the observation from inclusive jet analysis (suggestive of jet broadening from $R = 0.2$ to $R = 0.4$) there appears to be a smooth jet broadening trend. Note that the recoil jet $\pT$ spectrum is much flatter (harder) than the inclusive one, therefore impact of uncertainties on background fluctuations is much reduced.

Full jet reconstruction is not feasible for $R > 0.4$ due to large background fluctuations. To investigate the jet broadening on the away side of the trigger jet, jet-hadron (JH) azimuthal correlations between trigger jet and charged hadrons (detected by the TPC detector) are measured~\cite{joern}. The raw azimuthal correlation is parametrized via two Gaussian peaks (near and away side jet) and $\flow$ modulated background (with fixed $v_{2}$ values). The uncertainties in the (a priori unknown) jet $\flow$ value were chosen to cover the extreme cases of no $\flow$ and $50\%$ higher than $\flow\{2\}$ at $\pT = 6~\gevc$ (default is $\flow\{2\}$ at $\pT = 6~\gevc$). The associated track $\flow$ values and uncertainties follow the analysis in~\cite{v2_corr}. Due to ambiguities of ZYAM for broad jet structures, the background level was determined by the fit. For comparison ZYAM was applied (as expected for broad structures it leads to an underestimation of the correlated away-side yields for lower associated $\pT$).

\begin{figure}[htb]
\begin{minipage}[h]{0.48\textwidth}
\includegraphics[width=\textwidth]{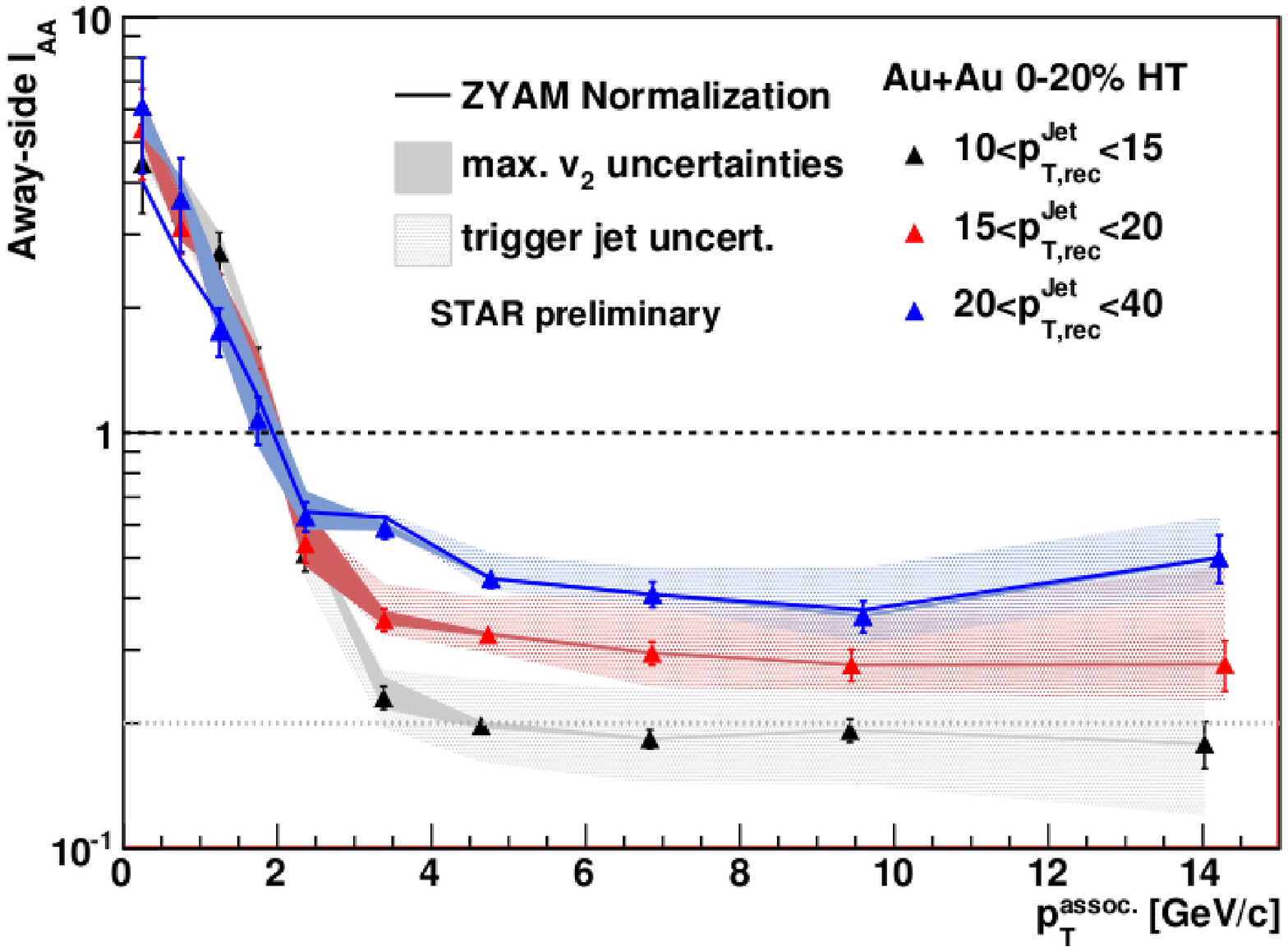}
\vspace{-10mm}
\caption{\label{fig:ASIaa}Away side $\iaa$ in JH correlations~\cite{joern}.}
\end{minipage}
\hfill
\begin{minipage}[h]{0.48\textwidth}    
\includegraphics[width=\textwidth]{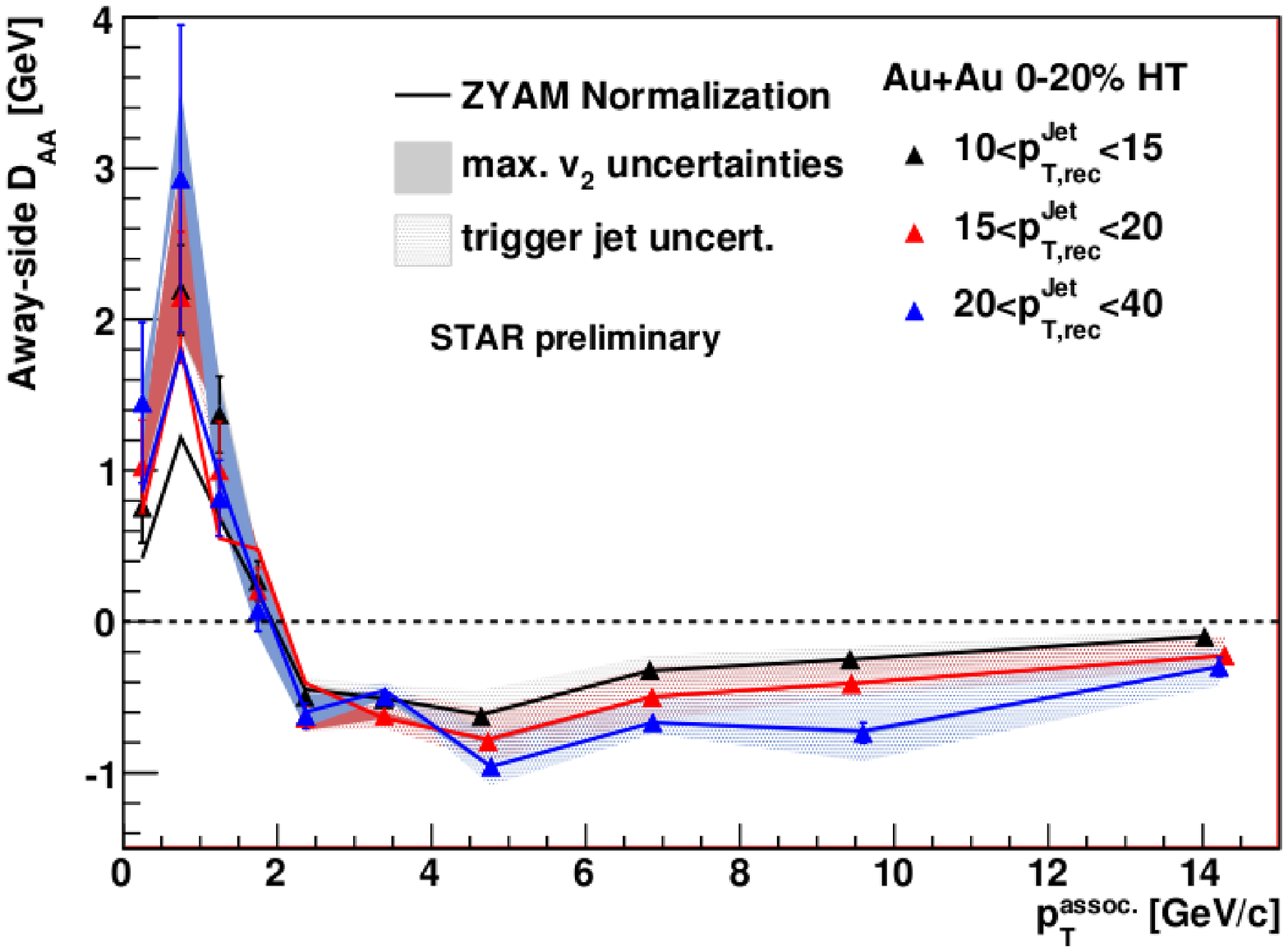}
\vspace{-10mm}
\caption{\label{fig:ASDaa}Away side $\daa$ in JH correlations~\cite{joern}.}
\end{minipage}
\end{figure}

Figure~\ref{fig:ASwidth} shows the awayside Gaussian width of JH in p+p and $0-20\%$ most central Au+Au collisions. There is a significant broadening (Au+Au w.r.t. p+p) for $\pTass<3~\gevc$, while no broadening at higher $\pTass$ is observed. $\iaa$, the ratio of per-trigger associated yields, is plotted in Figure~\ref{fig:ASIaa}. There is a significant suppression of high $\pT$ particles on the away side accompanied by an enhancement at low $\pTass$. In order to quantify the energy redistribution on the away side, it's better to instead of $\iaa$ use $\daa$:

\begin{equation}
\daa(\pTass) = \pTass \cdot(Y_\mathrm{AA}(\pTass) - Y_\mathrm{pp}(\pTass)), 
\end{equation}

where $Y_\mathrm{AA,pp}$ are per-trigger associated yields in AA,pp. Away side $\daa$ for JH is shown in Figure~\ref{fig:ASDaa}. In fact, the energy ``lost'' at high $\pT$ is approximately compensated by low $\pT$ enhancement~\cite{Alice_WWND}: jet quenching in action.

Given the observed broadening for $\pTass < 3~\gevc$, no broadening for high $\pTass$ and $\iaa$ shape independent of $\pTass$ at high $\pTass$, one can speculate that the original parton loses energy by emission of soft radiation (and therefore the original jet direction changes little: no broadening is observed at high $\pTass$). These soft fragments traverse the medium, receive transverse kicks and therefore appear at large angles with respect to the original parton direction. The energy loss is followed by a possibly vacuum-like fragmentation of a parton with reduced energy.

\vspace{-5mm}

\vspace{-5mm}

\vspace{-5mm}

\section{Summary}

We have presented STAR results on full jet reconstruction in d+Au collisions. Within current systematics there appears to be binary collision scaling compared to p+p collisions, but the final measurements (jet $\raa$ and $\rcp$) with reduced systematic uncertainties are yet to be completed.
The study of background fluctuations $\delta\pT$ in Au+Au collisions suggests its independence of $\pT^\mathrm{embed}$ and the probe jet fragmentation pattern. The shape of $\delta\pT$ will be used to unfold the measured jet $\pT$ spectrum to obtain the final result with decrased systematic uncertainties. 
The hints of jet broadening obtained first in inclusive jet $\pT$ spectrum and di-jet correlations were further studied using jet-hadron correlations, where significant broadening and enhancement at low $\pTass$ was accompanied by suppression (and no broadening) at high $\pTass$. This is consistent with the picture of pQCD-like energy loss to soft fragments followed by (possibly vacuum-like) fragmentation of a parton with reduced energy.

\section*{Acknowledgement}
This work was supported in part by grants LC07048 and LA09013 of the Ministry of Education of the Czech Republic.

\end{document}